\documentclass[12pt,preprint]{aastex}

\def\la{\mathrel{\mathpalette\fun <}}
\def\ga{\mathrel{\mathpalette\fun >}}
\def\fun#1#2{\lower3.6pt\vbox{\baselineskip0pt\lineskip.9pt
  \ialign{$\mathsurround=0pt#1\hfil##\hfil$\crcr#2\crcr\sim\crcr}}}

\slugcomment{Submitted to Astrophysical Journal Letters}

\shorttitle{CDM predicts MOND}
\shortauthors{M. Kaplinghat \& M. Turner}

\begin{document}

\title{How Cold Dark Matter Theory Explains Milgrom's Law}

\author{Manoj Kaplinghat\altaffilmark{1} and
Michael Turner\altaffilmark{1,2,3}}
\altaffiltext{1}{Department of Astronomy and Astrophysics,
The University of Chicago, 5640 S. Ellis Ave., Chicago, IL 60637-1433, USA}
\altaffiltext{2}{NASA/Fermilab Astrophysics Center,
Fermi National Accelerator Laboratory, PO Box 500,
Batavia, IL  60510-0500, USA}
\altaffiltext{3}{Department of Physics, Enrico Fermi Institute,
The University of Chicago, Chicago, Illinois 60637-1433, USA}

\begin{abstract}

Milgrom noticed the remarkable fact that the gravitational effect
of dark matter in galaxies only becomes important where
accelerations are less than about $10^{-8}\,{\rm cm / s^2}
\sim cH_0$ (``Milgrom's Law'').  This forms the basis for Modified
Newtonian Dynamics (MOND), an alternative to particle dark
matter. However, any successful theory of galactic dynamics must account
for Milgrom's Law. We show how Milgrom's Law comes about in the Cold
Dark Matter (CDM) theory of structure formation.

\end{abstract}

\section{Introduction}

The dark-matter mystery has been with us since Zwicky noticed
that the gravitational action of luminous matter is not sufficient to hold
clusters together \citep{zwicky33,smith36}.  Rubin and others brought
the problem closer to home by showing that spiral galaxies
like ours suffer the same problem (see e.g., \citet{knapp87}).
While the leading explanation for the dark matter problem today is
slowly moving, weakly interacting ``nonluminous'' elementary particles
remaining from the earliest moments -- cold dark matter (see e.g.,
\citet{turner00}) -- there is still interest in the possibility that the
explanation involves new gravitational physics (see e.g.,
\citet{sellwood00}). It is important to realize that particle dark
matter does exist -- the SuperKamiokande evidence \citep{fukuda98} for
neutrino mass implies that neutrino dark matter accounts for as much, or
perhaps more, matter as do bright stars.

Any gravitational explanation must deal with the fact that
the shortfall of the Newtonian gravity of luminous matter
occurs at widely different length scales -- at distances
much less than 1 kpc in dwarf spirals to distances greater than
100 kpc in clusters of galaxies.
Merely strengthening gravity beyond a fixed distance cannot
explain away the need for dark matter.

In 1983 \citet{milgrom83a,milgrom83b} made a remarkable observation:
the need for the gravitational effect of nonluminous (dark)
matter in galaxies only arises when the
Newtonian acceleration is less than about
$a_0 = 2 \times 10^{-8}\,{\rm cm\,s^{-2}} = 0.3\,cH_0$.
(Here, $H_0 = 70\pm 7\, {\rm km\,s^{-1}\, Mpc^{-1}}
= 100h\, {\rm km\,s^{-1}\, Mpc^{-1}}$ is present expansion rate of the
Universe.) This fact, which we will refer to as Milgrom's law, is the
foundation for his Modified Newtonian Dynamics (MOND) alternative to
particle dark matter. It is not our claim that the analysis to follow
rules out MOND.  


The correctness or incorrectness of MOND aside, the empirical fact that
the need for dark matter in galaxies always seems to occur at an
acceleration of around $cH_0$  must be explained by a successful theory
of structure formation. This {\em Letter} shows how Milgrom's Law arises
in the cold dark matter theory of structure formation. 

\section{How CDM Predicts Milgrom's Law}
\subsection{CDM theory}

The cold dark matter theory of structure formation has two basic
features:  seed density inhomogeneity that arose from quantum
fluctuations during inflation, and dark matter existing
in the form of slowly moving particles left over from the big bang. The
two leading candidates for the CDM particle are the axion and the
neutralino.  Each is predicted by a compelling extension of the
standard model of particle physics motivated by particle-physics
considerations (rather than cosmological) and has a predicted
relic density comparable to that of the known matter
density (see e.g., \citet{turner00}).

A recent estimate of the matter density puts the total at
$\Omega_M = 0.330 \pm 0.035$ and baryons at
$\Omega_B = 0.040\pm 0.008$ \citep{turner01}.
This means that CDM particles contribute
$\Omega_{\rm CDM} = 0.29\pm 0.04$ (less the contribution
of neutrinos).  (\citet{croft99} argue based upon the
formation of small-scale structure, that neutrinos can contribute no
more than about 10\% of the critical density.)

For our purposes here, a less essential feature of CDM is the
fact that the bulk of the critical density exists in the form of
a mysterious dark energy ($\Omega_X \simeq 0.66\pm 0.06$;
see e.g., \citet{turner01}).
While the existence of dark energy affects the details of structure
formation enough so that observations
can discriminate between a matter-dominated flat Universe and
one with dark energy, for the purposes of showing how CDM predicts
Milgrom's Law, dark energy and its character are not critical.
This is because most galaxies formed while the Universe was still
matter-dominated and well described by the Einstein -- deSitter model.

In the CDM scenario, structure forms from the bottom up,
through hierarchical merging of small halos to form larger
halos (see e.g., \citet{blumenthal84}).
The bulk of galactic halos formed around redshifts
of 1 to 5, with clusters forming at redshifts of 1 or less,
and superclusters forming today.  Within halos, baryons lose
energy through electromagnetic interactions and sink to the center,
supported by their angular momentum. Until baryonic dissipation
occurs, baryons and CDM particles exist in a universal ratio
of $\Omega_{\rm CDM}/\Omega_B \simeq 7$.  Were it not for
the concentration of baryons caused by dissipation, the gravity
of dark matter would be dominant everywhere.

\subsection{CDM and Milgrom's Law}

The CDM explanation for the gravitational effect of dark matter
``kicking in'' at a fixed acceleration approximately equal to $cH_0$
involves three ingredients:  i) the
fact that the Universe is reasonably well described by
the Einstein -- deSitter model during the period when
galaxies form; ii) the scale-free character
of the seed density perturbations over the relevant scales;
iii) baryonic dissipation; and iv) numerical coincidences.

The argument begins with facts i) and ii), which lead to the CDM
prediction of self-similar dark-matter halos. Halos, regardless of their
mass, can be described by the same mathematical form \citep[henceforth
NFW]{nfw}.  The exact functional form is not essential (see below);
for simplicity we write the halo profile for an object that began from
perturbations of comoving length scale $L$ as
\begin{equation}
\rho_L(r) \simeq \beta^3 \Omega_M \rho_{\rm crit} (1+z_c)^3 (r/{\ell})^{-2}
= \beta \Omega_M \rho_{\rm crit} (1+z_c) (r/L)^{-2}\,,
\label{eq:haloprofile}
\end{equation}
where $\rho_{\rm crit} = 3H_0^2/8\pi G$ is the critical density today,
$z_c$ is the redshift of halo collapse and $\beta$ is a numerical
constant of ${\cal O} (5)$.  The physical size of the perturbation after
collapse ($\equiv {\ell}$) is related to its comoving size, ${\ell} =
L/\beta (1+z_c)$; the factor of $1/(1+z_c)$ is due to the expansion of
the Universe and the factor of $1/\beta$ is due to collisionless
collapse. Because $\Omega_M (1+z_c)^3\rho_{\rm crit}$ is the mean matter
density at the redshift of collapse, Eq. \ref{eq:haloprofile} says that
the mean density of the collapsed structure interior to $r=\ell$ is
about 100 times the ambient density when collapse occurred.

The redshift of collapse is determined by the spectrum of density
perturbations:  collapse on length scale $L$ occurs when the {\em rms}
mass fluctuation on that scale ($\equiv \sigma_L$) is of order unity.
Neglecting nonlinear effects, $\sigma_L$ at redshift $z$ is related to
the matter power spectrum today  ($\equiv |\delta_k|^2$):
\begin{equation}
\sigma_L (z) = \left[ \int_0^\infty\,
 {k^2 |\delta_k|^2\over 2\pi^2}|W_L(k)|^2 dk \right]^{1/2}
 \simeq  (\epsilon /10^{-5} ) (1+z)^{-1}
 (L/L_0)^{-{1\over 2}(n_{\rm eff} +3)} \,,
\label{eq:collapse}
\end{equation}
where $k\sim L^{-1}$, $W_L(k)$ is the Fourier transform
of the top-hat window function, and $n_{\rm eff} \approx -2.2$ is the
logarithmic slope of $L^3\sigma_L^2 \sim |\delta_k|^2$ (with respect to
$k$) around galaxy scales.\footnote{For exactly scale-invariant density
perturbations, $n_{\rm eff}$ varies from $-2.5$ to $-2$ for
$L=0.01\,$Mpc to $L=1\,$Mpc in SCDM.  Inflation does not predict
precisely scale-invariant density perturbations (see e.g.,
\citet{huterer00}). In the case of nonscale-invariant density
perturbations, $n_{\rm eff} = -2.2 + (n-1)$, where $n-1$ quantifies the
deviation from scale invariance and is expected to be of order $\pm
0.1$.} The quantity $\epsilon$ is the  dimensionless amplitude of the
primeval fluctuations in the gravitational potential, determined by COBE
to be about $10^{-5}$, and $L_0\simeq 10h^{-1}\,$Mpc is
the scale of nonlinearity today (for $\epsilon \sim 10^{-5}$).
Substituting Eq. \ref{eq:collapse} into Eq. \ref{eq:haloprofile}, it
follows that
\begin{equation}
\rho_L(r) = [(3\beta/8\pi)\, \Omega_M\, (\epsilon /10^{-5})]\,
(H_0^2/G)\, (L/L_0)^{-{1\over 2}(n_{\rm eff} +3)}\, (r/L)^{-2} \,.
\label{eq:haloprofile2}
\end{equation}

The third ingredient is baryonic dissipation: after halos form,
their baryons dissipate energy and collapse in linear scale by
a factor $\alpha \approx 10$ to form a disk supported by angular
momentum (see e.g., \citet{dalcanton97}). The degree of baryonic
collapse is determined by the dimensionless spin parameter $\lambda$,
which is the ratio of the angular velocity of the galaxy to the angular
velocity that would be required to support the structure purely by
rotation. The angular momentum of galaxies is thought to arise from
tidal torquing \citep[and references therein]{peebles69}. Theory and
simulations \citep{warren92} seem to agree that $\lambda$ is independent
of scale, with a median value of $\lambda \approx 0.05$. If one
assumes that the angular momentum of the gas is conserved during disk
formation, then (see \citet{padmanabhan})
$\alpha \approx \Omega_B \lambda_D^2/\Omega_M \lambda^2$, which is about
12 because the disk spin parameter $\lambda_D \approx 0.5$.

Because of the increased concentration of baryons interior to
$r\sim {\ell}\, (\Omega_{\rm B}/\Omega_{\rm CDM})$,
their gravity will dominate the dynamics in the inner
regions. (This statement is true as long as
$\alpha > \Omega_{\rm CDM}/\Omega_{\rm B}$.) 
Thus, the transition from dark-matter dominated gravity to
luminous-matter dominated gravity should occur around
$r_{\rm DM} = L/7 \beta (1+z_c)$. The acceleration at the point when
dark matter gravity begins to dominate is
\begin{eqnarray*}
 a_{\rm DM} \equiv a(r_{\rm DM}) =
 	{GM(r_{\rm DM})\over r_{\rm DM}^2} = [4\pi G\, ({\ell}/7)]
 	\rho_L({\ell}/7) \,.
\end{eqnarray*}
After some re-writing, Milgrom's Law emerges
\begin{eqnarray}
a_{\rm DM} & = & cH_0 \left[ 10\beta^2 \,\Omega_M
   \left( {\epsilon \over 10^{-5}}\right)^2 \right]
     (c^{-1}H_0L_0) \left( {L\over L_0}\right)^{-n_{\rm eff} -2} \,,
\nonumber\\
     & = & {\cal O}(1)\, cH_0 \left( {L\over L_0}\right)^{0.2} \,.
\label{eq:adm}
\end{eqnarray}
The final ingredient is the conspiracy of numerical factors to
give a coefficient of unity and a very mild scale dependence (over
3 orders of magnitude in mass, $a_{\rm DM}$ changes by only
a factor of 1.6).

We have assumed in the above discussion that most of the baryons in the
protogalaxy dissipate and form disks. How valid is this assumption?
Clearly some of the baryons will be inhibited from collapsing by the UV
radiation field, or blown away into the inter-galactic medium due to
feedback from supernovae and possibly other phenomena related to star
formation. It is sensible to assume that these effects are more
pronounced for smaller mass galaxies. However, so long as the fraction
of baryonic matter that collapses does not vary strongly with scale our
analysis goes through with only numerical factors changing. In fact, if
the collapsed fraction in a $0.1 {\cal L}_\star$ galaxy (${\cal L}$
denotes the luminosity) is about half that of a ${\cal L}_\star$ galaxy,
this would give rise to a scale dependence in $a_{\rm DM}$ about the
same as, but {\em opposing} the change, in Equation \ref{eq:adm} --
leaving $a_{\rm DM}$ essentially scale-free. Of course, at the low mass
end of the galactic scale, one could have a much smaller collapsed
fraction that could introduce scatter or deviation in the $a_{\rm DM}$ vs
luminosity relation (even after taking into account the fact that on
those small scales $n_{\rm eff}$ is smaller than -2.2). In this case, an
accurate derivation of Milgrom's law will require more sophisticated
models incorporating gas dynamics of the baryons.   


The mild scale dependence of the acceleration where dark matter
dominates owes to the fact that $n_{\rm eff} \approx -2$, around
galactic scales.  It arises from a combination of the primeval spectral
index ($n\simeq 1$) and the bending of the shape of the spectrum of
perturbations caused by the fact that perturbations on small scales ($k
\ga 0.1\,{\rm Mpc^{-1}}$) entered the horizon when the Universe was
radiation-dominated and those on large scales ($k\la 0.1\,{\rm
Mpc^{-1}}$) entered the horizon when the Universe was
matter-dominated. For $k\ll 0.1\,{\rm Mpc^{-1}}$, 
$n_{\rm eff}\rightarrow 1$ and for $k\gg 0.1\,
{\rm Mpc^{-1}}$, $n_{\rm eff}\rightarrow -3$. 

Returning to the numerical conspiracy that leads to
$a_{\rm DM}\approx cH_0$; for $n_{\rm eff} = -2$, the factor
$(\epsilon /10^{-5})^2 L_0$ is just the scale of nonlinearity today,
independent of the actual value of $\epsilon$.  The numerical
coincidence then is the fact that the scale of nonlinearity today is
much less than the Hubble scale.  \citet{scott01} have tied this fact to
the cooling scale of baryons, which can be related to fundamental
constants and $\epsilon$.

Equation \ref{eq:adm} only holds around galaxy scales ($L\sim 1\,$Mpc),
where $n_{\rm eff} \approx -2 $ and $\alpha \sim 10$.
Clusters are dark-matter dominated almost everywhere because cluster
baryons do not dissipate significantly. Milgrom's law would, therefore,
assert that the Newtonian acceleration in clusters should be less
than $cH_0$ almost everywhere -- in contradiction with observations.
Said another way, CDM correctly predicts that Milgrom's Law {\em
should not} apply to clusters.

The issue of the shape of the halo density profile is not central to our
arguments.  We have repeated our calculation for
the NFW profile and find
$a_{\rm DM}\approx 10^{-8}(M_{200}/10^{12}M_\odot)^{0.1}\,{\rm cm\,s^{-2}}$,
which is similar to the result obtained in Eq. \ref{eq:adm},
($M_{200}$ is the mass interior to the point where the density is
200 times the critical density).
MOND automatically predicts asymptotically flat
rotation curves; in CDM the flatness of rotation curves has its origin
in the fact that over a significant portion of the halo,
$\rho_{\rm halo} \propto r^{-2}$.  The NFW halo profile asymptotes to
$\rho_{\rm halo} \rightarrow r^{-3}$ so that CDM predicts
$v_{\rm circular} \rightarrow \sqrt{\ln r /r}$.

Another coincidence for CDM is known. The galaxy-galaxy correlation
function is very well fit by a power law, $\xi(r)= (r/r_0)^{-1.8}$ where
$r_0 = 5h^{-1}\,$Mpc (see e.g., \citet{groth77,baugh96}).
In CDM theory, the two-point correlation function of mass is not a good
power law; however, when bias is taken into account (the nontrivial
relation between mass and light), the galaxy -- galaxy correlation
function turns into a power-law (see e.g., \citet{pearce99}), in good
agreement with observations.

\section{Concluding remarks}

The derivation of Eq. \ref{eq:adm} is the key result of this paper. It
illustrates how Milgrom's Law -- the need for dark matter in galaxies at
accelerations less than about $cH_0$ -- arises in CDM theory.
While scale-free density perturbations, an epoch where the Universe
is well described by the Einstein -- deSitter model and
baryonic dissipation are essential, the fact that $a_{\rm DM}$
is nearly $cH_0$ appears to be a numerical
coincidence.  Furthermore, $a_{\rm DM}$ is a fixed number since
galaxies are bound and well relaxed today, while $cH$ decreases with
time.  Thus, the approximate equality of $a_{\rm DM}$ with $cH$ only
holds today.

The purpose of our {\em Letter} was to illuminate the basic physics
that underlies the emergence of Milgrom's Law within CDM theory.
It was not our intention to present a detailed analysis.
To achieve our purpose
we made some strong -- but we believe reasonable -- assumptions.
The strongest of these assumptions is that all the
baryonic matter associated with galaxies dissipates and collapses.  This
is probably not true, as significant amounts of baryonic material still
exist in hot gas \citep{fukugita98}. However, so long as the fraction of
baryonic matter that collapses does not vary much with scale (which we have
quantified in the previous section) our analysis goes through
with only numerical factors changing.  If the fraction of baryonic
matter that collapses does vary dramatically with scale, one would
expect deviant objects and scatter.  We remind the reader that a
detailed, semi-analytic calculation of galactic rotation curves in CDM 
\citep{vandenbosch00} 
clearly shows the presence of a characteristic acceleration scale (of
the order of $cH_0$), 
and they fit the data about as well as those derived from MOND.
(We would argue that Eq. \ref{eq:adm} is the underlying explanation for
the appearance of this acceleration scale.) 
Further, a study of about 1000 spiral galaxies \citep{persic96}, with
luminosities from about $0.1 {\cal L}_\star$ to ${\cal L}_\star$,
is in agreement with $r_{\rm DM}/ \ell$ being approximately 
constant.\footnote{\citet{persic96} find a relationship
between two observationally based quantites: $R_T$, which is essentially
scales as our $r_{\rm DM}$, and ${\cal L}$.
Assuming $r_{\rm DM}/ \ell$ is constant and the observationally
inferred baryonic mass vs. luminosity relation \citep{salucci99}, we are
able to reproduce the $r_{\rm DM}$ vs. ${\cal L}$ scaling of
\citet{persic96}.   
}
We have verified that for these galaxies the variation in $a_{\rm DM}$
over the 1 order of magnitude range in luminosity is less than 20\%.
Both of these studies lend credence to our underlying assumptions.

Separating the important clues from the misleading coincidences is at
the heart of scientific creativity.  Hoyle's observation that the energy
released in burning $25\%$ of the Hydrogen to Helium is approximately
equal to that of the CMB suggested a non big-bang origin for the CMB
(see e.g., Burbidge \& Hoyle, 1998).   In the end, it turned out to be
a misleading coincidence.  Within the big-bang model, Holyle's
coincidence is explained by the near equality of
the dimensionless amplitude of density perturbations $\epsilon$ and
the product of the efficiency of nuclear burning times
$\Omega_B$:  To make stars by the present epoch, $\Omega_\gamma$
must be $\sim \epsilon $, which coincidentally is equal to the
energy that would be released in producing the observed Helium
abundance (Martin Rees, private communication).

CDM not only predicts Milgrom's Law (at least over a order of magnitude
range in luminosity from $0.1 {\cal L}_\star$ to ${\cal L}_\star$), but
also accounts for a wealth of other cosmological observations. This
suggests to us that Milgrom's Law is a misleading coincidence rather
than evidence for a modification of Newtonian dynamics.

\acknowledgments
This work was supported by the DOE (at Chicago and Fermilab),
the NASA (grant NAG5-10842 at Fermilab) and the NSF (grant PHY-0079251
at Chicago). We thank Martin Rees for stimulating conversations,
and Arthur Kosowsky, Douglas Scott and Frank van den Bosch
for useful comments.

\end{document}